\documentstyle[12pt,epsf]{article}
\begin{document}
\begin{titlepage}
\begin{flushright}
  KUNS-1838\\
\end{flushright}

\begin{center}
\vspace*{10mm}
  
{\LARGE \bf Twisted Moduli Stabilization \\ 
\vspace{1.5mm}
in  Type I String Models}
\vspace{12mm}

{\large
Tetsutaro~Higaki\footnote{E-mail address:
  tetsu@gauge.scphys.kyoto-u.ac.jp} ~and~ 
Tatsuo~Kobayashi\footnote{E-mail address:
  kobayash@gauge.scphys.kyoto-u.ac.jp}
}
\vspace{6mm}

{\it Department of Physics, Kyoto University,
Kyoto 606-8502, Japan}\\[1mm]

\vspace*{15mm}

\begin{abstract}
We consider the model with the dilaton and twisted moduli fields, 
which is inspired by type I string models.
Stabilization of their vacuum expectation values is studied.
We find the stabilization of the twisted moduli field 
has different aspects from the dilaton stabilization.
\end{abstract}

\end{center}
\end{titlepage}

\section{Introduction}

Superstring theory is the promising candidate for 
unified theory including gravity.
It has no free parameter and 
gauge couplings, Yukawa couplings and others of 
low energy effective field theory 
are determined by vacuum expectation values (VEVs) 
of dilaton/moduli fields.
Thus, it is important to stabilize these VEVs.
Indeed, several stabilization mechanisms have been proposed.

The dilaton/moduli fields have no potential perturbatively. 
Only non-perturbative effects lift their potential.
Gaugino condensations are non-perturbative effects, 
which are plausible to generate a non-perturbative superpotential 
of dilaton/moduli fields.
However, one can not stabilize the VEV of the dilaton field 
to a fine value in 
the model with a superpotential generated by 
a single gaugino condensation and the tree-level K\"ahler 
potential.
One of the simple extensions is the model with 
double gaugino condensations and the tree-level 
K\"ahler potential, i.e. the so-called 
racetrack model \cite{Krasnikov:jj}-\cite{Dine:1999dx}, 
while non-perturbative 
K\"ahler potential has also been 
considered \cite{Banks:1994sg,Binetruy:1996xj,Casas:1996zi}.
In fact, one can stabilize the VEV of the dilaton field to 
a finite value depending on beta function coefficients 
of gauge couplings relevant to gaugino condensations.

Twisted moduli fields appear in orbifold/orientifold models.
These are localized at fixed points.
In type I models, twisted moduli fields are gauge singlets, 
while they are charged in heterotic models.
Gauge kinetic functions depend on twisted moduli 
in type I models \cite{Aldazabal:1998mr,Ibanez:1998rf}.
They play a role in 4D Green-Schwarz anomaly cancellation 
e.g. for anomalous $U(1)$\cite{Ibanez:1998qp,Lalak:1999bk}, 
while the dilaton field plays the same role 
in heterotic models \cite{DSW}.\footnote{See also Ref.\cite{KN} 
for anomalous U(1) in heterotic models.}
Thus, their VEVs determine the magnitude of Fayet-Iliopoulos 
terms.
The prediction of the gauge couplings depends on the VEV of twisted 
moduli fields.
The mirage unification of gauge couplings is 
one possibility to explain the experimental 
values of gauge couplings with lower string scale \cite{Ibanez:1999st}.
Hence, the magnitude of twisted moduli field VEVs is 
phenomenologically important.

In this paper, we will consider the model with dilaton and twisted 
moduli fields, which is inspired by type I string models, and 
study stabilization of dilaton 
and twisted moduli fields.
For similar purpose, models with twisted moduli fields 
have been studied in Refs.~\cite{Abel:2000tf,Ciesielski:2002fs}.
The K\"ahler potential of the twisted moduli fields 
is not clear.
Here we will use the assumption of the canonical form, 
which has been studied in 
Ref.~\cite{Poppitz:1998dj}\footnote{See also 
Ref.~\cite{Scrucca:2000za}.} and show this form is important 
to stabilize the VEV of twisted moduli.
As another example, we will assume the  logarithmic form
of the K\"ahler potential  for the twisted moduli fields 
like the dilaton and other moduli fields.
That is an example of K\"ahler potentials, 
which have a different behavior from the canonical form.
However, we will show that even in the case with 
the logarithmic K\"ahler potential the positive exponent 
in the non-perturbative superpotential is useful  
for the stabilization of the twisted moduli fields.

This paper is organized as follows.
In the next section, we review shortly the stabilization 
to the dilaton VEV in the racetrack model.
In section 3, we study the model with 
dilaton and twisted moduli fields.
In section 3.1 we shortly review about the twisted moduli fields.
In section 3.2 we consider the single gaugino condensation 
model and show how different the stabilization of 
twisted moduli fields is from 
the dilaton stabilization.
In section 3.3 we consider a specific double gaugino condensation 
model in order to study the simultaneous stabilization of 
the dilaton and twisted moduli fields.
In section 3.4 we give a comment about effects of 
twisted moduli fields on the dilaton VEV.
Seciton 4 is devoted to conclusion and discussions.

\section{The racetrack model}

The tree-level K\"ahler potential of the dilaton field 
is obtained as 
\begin{equation}
K = - \ln (S + \bar S).
\label{Kahler-S}
\end{equation}
The gauge kinetic function of heterotic models 
is obtained as 
\begin{equation}
f=S,
\end{equation}
up to Kac-Moody level, and the gauge coupling $g$ is 
obtained as $Re(S) = 1/g^2$.
This is the same for the gauge multiplets originated from 
$D9$ branes in type I models.
Perturbatively, the dilaton field has a flat potential.
The single gaugino condensation induces 
the non-perturbative superpotential,
\begin{equation}
W = d e ^{-\Delta S},
\label{superW-1}
\end{equation}
where $d$ is a constant, $\Delta = -24\pi^2/b$ and $b$ is the 
one-loop beta 
function coefficient, e.g. $b= -3N_c$ for 
pure N=1 $SU(N_c)$ Yang-Milles theory.
With the above K\"ahler potential, the scalar potential $V$ 
is written as
\begin{equation}
V = \frac{1}{S + \bar S}\left[|(S+\bar S)W_S-W|^2-3|W|^2\right],
\label{scalar-V}
\end{equation}
where $W_S$ denotes the first derivative of $W$ by $S$, 
i.e. $W_S = \frac{\partial W}{\partial S}$. 
Here we have not taken into account $D$-terms, 
although $S$ has a $D$-term potential in heterotic models if 
the model has anomalous $U(1)$.\footnote{In this case, 
the dilaton field $S$ is relevant to 
Green-Schwarz anomaly cancellation.}
We have the following solutions of $\partial V/\partial S =0$ :
\begin{equation}
(S+\bar S)W_S-W =0,
\label{sol-1}
\end{equation}
or 
\begin{equation}
(S+\bar S)^2W_{SS} = 2 \bar W \frac{(S+\bar S)W_S-W }
{(S+\bar S)\bar W_S-\bar W } .
\label{sol-2}
\end{equation}

With the single gaugino condensation superpotential (\ref{superW-1}), 
the solution (\ref{sol-1}) leads to $S+\bar S= -\frac{1}{\Delta}$, 
which is not a realistic VEV for $S$ in the asymptotically free case.
The solution (\ref{sol-2}) leads to $\Delta(S+\bar S) =\sqrt{2}$, 
but this corresponds to the maximum point of $V$.
See Fig.~1, where the lower line shows the 
scalar potential against $s\equiv S + \bar S $ in the case 
with $\Delta =10$ and $d=1$.

In heterotic models, the requirement of 
$SL(2,Z)$ duality invariance of the overall moduli field $T$ 
leads to the following 
superpotential \cite{Ferrara:1989bc,Ferrara:1990ei,Font:1990nt},
\begin{equation}
W= d e^{-\Delta S}\hat W(T).
\end{equation}
The corresponding scalar potential is written as 
\begin{equation}
V= \frac{|e^{-\Delta S}|^2 |\hat W(T)|^2}{(S + \bar S)(T+\bar T)^3}
[((S+\bar S)\Delta +1)^2 + g(T,\bar T)],
\end{equation}
with
\begin{equation}
g(T,\bar T) \equiv \frac{1}{3}|(T+\bar T)\frac{W_T}{W}-3|^2 -3.
\end{equation}
Here we have used the K\"ahler potential of $T$ as 
\begin{equation}
-3 \ln(T + \bar T) .
\end{equation}
However, the inclusion of $\hat W(T)$ does not help 
the stabilization of $S$.
If $g(T,\hat T)< -1$, the situation is the same as the case 
without $\hat W(T)$.
If $g(T,\hat T)> -1$, the scalar potential monotonically decreases 
as $s$.
The upper line in Fig.~1 shows $(T+\bar T)^3V/|\hat W(T)|^2$ for 
$\Delta = 10$, $d=1$ and $g(T,\bar T) =-0.5$.

\begin{figure}
\epsfxsize=0.7\textwidth
\centerline{\epsfbox{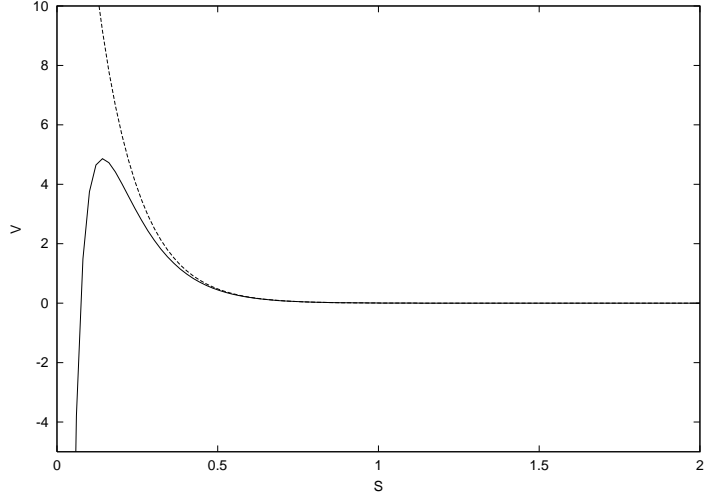}}
\caption{The lower line corresponds to the scalar potential 
for $\Delta = 10$ and $d=1$ without $\hat W(T)$.
The upper line corresponds to $(T+\bar T)^3V/|\hat W(T)|^2$ for 
$\Delta = 10$, $d=1$ and $g(T,\bar T) =-0.5$.}
\label{fig1}
\end{figure}

One mechanism to stabilize the VEV of $S$ is to consider 
the superpotential with double gaugino condensations, 
\begin{equation}
W = d_1 e^{-\Delta_1 S} + d_2 e^{-\Delta_2 S} .
\end{equation}
With this superpotential, the solution (\ref{sol-1}) of 
$\partial V /\partial S =0$ is given as 
\begin{eqnarray}
Im(S) &=& \frac{\pi}{\Delta_1 -\Delta_2}(2n+1) ,\\
Re(S) &=& \frac{1}{\Delta_1 -\Delta_2} 
\ln \frac{(1+2\Delta_1 Re(S))d_1}{(1+2\Delta_2 Re(S))d_2} .
\label{RT-sol-1}
\end{eqnarray}
If $\Delta_a Re(S) \gg 1$, the latter equation becomes 
the simple equation,
\begin{equation}
Re(S) = \frac{1}{\Delta_1 -\Delta_2} 
\ln \frac{\Delta_1 d_1}{\Delta_2d_2} .
\label{RT-sol-2}
\end{equation}
In the case with $\ln \frac{\Delta_1 d_1}{\Delta_2d_2}=O(1)$, 
the stabilized value of $Re(S)$ is determined by 
$1/(\Delta_1 -\Delta_2)$.
Thus, the natural order of $S$ is of $O(1/\Delta)$.
If $\Delta_1$ and $\Delta_2$ are close each other, 
the VEV of $S$ is enhanced.
For example, one can obtain $O(1)$ of $S$ in two case; 
i)  the case with large beta-function coefficients\footnote{
In Ref.~\cite{Kaplunovsky:1997cy} large beta-function coefficients 
are studied from the viewpoint of F-theory.}
$b_i=O(100)$ and ii) the case with fine-tuning values  
($\Delta_1, \Delta_2$).
For the latter case, we have $Re(S)=1$ 
e.g. for  $(\Delta_2-\Delta_1)/\Delta_1 = 0.04$ 
and $b_1=10$.

\section{Stabilization in model with twisted moduli}

\subsection{Twisted moduli}

Twisted moduli fields $M$ are localized 
at orbifold fixed points and these 
moduli fields are important from several phenomenological 
view points in 4D models obtained from type I 
and type II orientifold models.
For example, the gauge kinetic functions 
corresponding to gauge groups originated from $D9$ branes 
are written as
\begin{equation}
f_a = S + \sigma_a M,
\end{equation}
where $\sigma_a$ is a model-dependent 
constant \cite{Ibanez:1998qp,Antoniadis:1999ge}.
Concerned about $\sigma_a$, here we take purely phenomenological 
standpoint, that is, we treat $\sigma_a$ as free parameters.  
Similarly, for the gauge groups originated from e.g. 
$D5$-branes, which are wrapped on the $i$-th torus ($i=1,2,3$), 
the corresponding gauge kinetic functions are written as 
\begin{equation}
f_{5a} = T_i + \sigma_{5a} M,
\end{equation}
where $T_i$ is the moduli field corresponding to the
$i$-th torus and its K\"ahler potential is obtained 
\begin{equation} 
K(T_i,\bar T_i)   = -\ln(T_i + \bar T_i),
\end{equation}
that is, its form is exactly the same as the K\"ahler potential of 
the dilaton field (\ref{Kahler-S}).
Thus, we can discuss the stabilization of $T_i$ due to 
gaugino condensation from $D5$-originated gauge groups in the same way 
as the stabilization of $S$ due to condensation from 
$D9$ gaugino fields.
Here, we concentrate ourselves on the $S$ stabilization.

One of important aspects is that 
the twisted moduli field $M$ plays a role in 
4D Green-Schwarz anomaly cancellation mechanism.
For example, under anomalous $U(1)$ symmetry, 
the twisted moduli fields are assumed to transform at one-loop level
\begin{equation}
M \rightarrow M +i \delta_{GS}\Lambda ,
\end{equation}
with the transformation parameter $\Lambda$.
The Fayet-Iliopoulos  term 
is written by the first derivative of K\"ahler potential 
$\frac{\partial K(M,\bar M)}{\partial M}$,
where $K(M,\bar M)$ is the K\"ahler potential of 
the twisted moduli field.
Thus, the magnitude of the Fayet-Iliopoulos  term is 
determined by the VEV of $M$.

Unfortunately, the K\"ahler potential of $M$, $K(M,\bar M)$, is 
still unclear.
In the limit $M\rightarrow 0$, the K\"ahler metric 
has  no singularity.
Hence, the K\"ahler potential $K(M,\bar M)$ could be expanded as 
\begin{equation}
 K(M,\bar M) =\frac{1}{2}(M +\bar M)^2 +\cdots .
\end{equation}
Actually, this form has been studied 
in Ref.~\cite{Poppitz:1998dj}.
Thus, we use the assumption of the K\"ahler potential as 
$K(M,\bar M) =\frac{1}{2}(M +\bar M)^2$ in a half part of our 
analyses.
However, since its reliability for $M=O(1)$ may be unclear, 
we assume \footnote{We would like to thank Kiwoon Choi 
for suggesting this point.}
\begin{equation}
 K(M,\bar M) =-\ln (M +\bar M) ,
\end{equation}
as a trial form of the K\"ahler potential for $M\geq O(1)$.

We also give comments on the gauge coupling unification.
Within the framework of minimal supersymmetric standard model (MSSM), 
three gauge couplings of $SU(3) \times SU(2) \times U(1)_Y$ 
meet around $M_X = 2\times 10^{16}$ GeV.
Suppose that the three gauge groups are originated from 
different sets of $D9$-branes.
If one can stabilize $Re(S) \gg \sigma_a Re(M)$, 
the gauge couplings are universal at the string scale $M_s$.
That implies $M_s \approx M_X$.
Otherwise, if $\sigma_a Re(M)$ is sizable, 
the gauge couplings are, in general, non-universal at $M_s$.
However, one of interesting possibilities to explain 
the experimental values of gauge couplings is the 
so-called "mirage unification" \cite{Ibanez:1999st}.
The MSSM gauge coupling at $\mu$ is obtained as 
\begin{equation}
\frac{1}{g^2_a(\mu)} = S + \sigma^{MSSM}_a M + 
\frac{b^{MSSM}_a}{16\pi^2} \ln \frac{M^2_s}{\mu^2} ,
\end{equation}
where $b^{MSSM}_a$ are the one-loop beta-function coefficient 
for the MSSM.
Let us consider a specific model that the constants 
$\sigma^{MSSM}_a$ 
are proportional to $b^{MSSM}_a$.
In this scenario, the gauge couplings are 
non-universal at $M_s$, but its prediction 
is the same as the universal gauge coupling around $M_X$.
The string scale $M_s$ can be low depending on 
$\sigma^{MSSM}_a M$.
Note that even a small value of $Re(M)$ like 
$\sigma^{MSSM}_a M =O(0.01)$ is important.
If the ratio of $M_s$ to $M_X$ satisfies  
\begin{equation}
\log_{10}\frac{M_s}{M_X} \sim \frac{\sigma^{MSSM}_a Re(M)}{0.03} ,
\end{equation}
that leads to the MSSM gauge couplings consistent with 
the experimental values.

Thus, it is important to study the stabilization of 
the twisted moduli field $M$.
That is the issue we will study in the flowing sections.
We will also discuss how the twisted moduli field $M$ 
affects on the stabilization of the dilaton field $S$.

\subsection{Single gaugino condensation}

Here we study the case with single gaugino condensation, 
although one can not 
stabilize the dilaton field with the single gaugino 
condensation as seen section 2.
That will be useful for later discussions.
The K\"ahler potential is written as 
\begin{equation}K = -\ln(S +\bar S) 
+K(M,\bar M),
\end{equation}
and the superpotential due to the gaugino condensation 
is obtained as
\begin{equation}
W= d e^{-\Delta (S + \sigma M)} .
\end{equation}
Using the K\"ahler potential and the superpotential, 
we can write the scalar potential as
\begin{eqnarray}
V &=& \frac{e^{K(M,\bar M)}}{S + \bar S}
[(K^{-1})^M_{\bar M}|\frac{\partial K(M,\bar M)}{\partial M}W - 
W_M|^2  \nonumber \\ 
&+&  |(S+\bar S)W_S-W|^2-3|W|^2 ], 
\label{scalar-V-M}
\end{eqnarray}
where $(K^{-1})^M_{\bar M}$ denotes the inverse of 
the K\"ahler metric for $M$ and $\bar M$.
Again, we do not take into account $D$-terms.
Inclusion of $D$-terms would be studied elsewhere.
For this scalar potential, one of solutions to 
the stationary condition $\frac{\partial V}{\partial M} =0$ is 
\begin{equation}
\left(\frac{\partial K(M,\bar M)}{\partial M}- \Delta \sigma \right)~W 
=0,
\label{sol-M}
\end{equation}
that is, $\frac{\partial K(M,\bar M)}{\partial M} =  \Delta \sigma$ 
is one solution.

\subsubsection{The case with $K=\frac{1}{2}(M+\bar M)^2$}

To be concrete, we use the assumption of 
the K\"ahler potential $K=\frac{1}{2}(M+\bar M)^2$.
In this case, it is convenient to define $m$ as 
\begin{equation}
m \equiv M+\bar M -\Delta \sigma .
\end{equation}
Then the scalar potential is written as 
\begin{equation}
V = \frac{e^{-\Delta(S + \bar S)-\sigma^2 \Delta^2/2}}{S + \bar S}
e^{m^2/2}[m^2+ g(S+\bar S)] ,
\end{equation}
where 
\begin{equation}
g(S+\bar S) \equiv | (S + \bar S)\frac{W_S}{W}-1|^2 -3 .
\label{def-g-0}
\end{equation}
For the single gaugino condensation, we have 
\begin{equation}
g(S+\bar S) = (\Delta (S + \bar S)+1)^2 -3 .
\label{def-g}
\end{equation}
The solutions of the stationary condition 
$\frac{\partial V}{\partial m}=0$ are obtained as follows,
\begin{equation}
m=0, \qquad m = \pm \sqrt{-2-g(S+\bar S)} .
\end{equation}
The former solution corresponds to Eq.(\ref{sol-M}).
The latter solutions are allowed only if 
\begin{equation}
2 + g(S +\bar S) < 0  .
\end{equation}
By the definition (\ref{def-g}), this 
inequality is never satisfied for $(S + \bar S) > 0$.
We have $\frac{\partial^2 V}{\partial m^2}  > 0$ for the former
solution $m=0$ if 
\begin{equation}
2 + g(S +\bar S) > 0  .
\end{equation}
By the definition (\ref{def-g}), this 
inequality is always satisfied for $(S + \bar S) > 0$.
In addition, for the latter solution, we always have 
$\frac{\partial^2 V}{\partial m^2}  > 0$ if the solution 
is realized, i.e. $2 + g(S +\bar S) < 0$.
In Fig.~2 the upper and lower lines show 
$v \equiv e^{m^2/2}[m^2+g(S+\bar S)]$ 
for $g(S+\bar S)=-1$ and $-3$, respectively.

\begin{figure}
\epsfxsize=0.7\textwidth
\centerline{\epsfbox{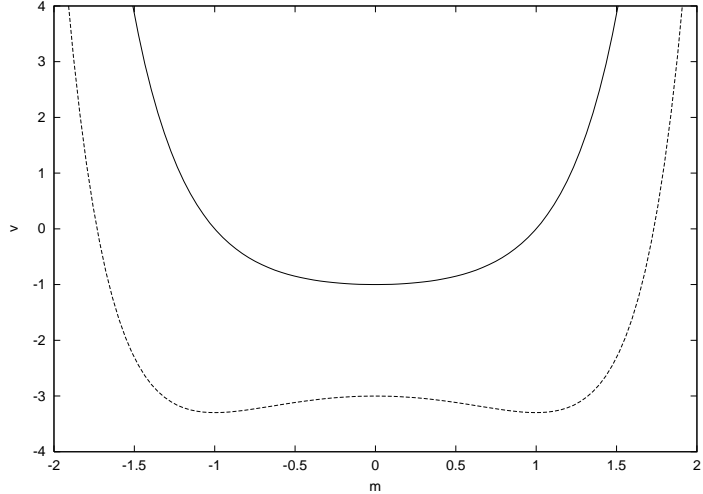}}
\caption{The upper and lower lines show 
$v \equiv e^{m^2/2}[m^2+g(S+\bar S)]$ 
for $g(S+\bar S)=-1$ and $-3$, respectively.}
\label{fig2}
\end{figure}

More explicitly, these solutions lead to the following values of 
$Re(M)$,
\begin{equation}
2Re(M) = -\Delta \sigma, \qquad 
2Re(M) = -\Delta \sigma \pm \sqrt{-2-g(S+\bar S)} .
\end{equation}
In the case that the K\"ahler potential 
$K(M,\bar M) = \frac{1}{2}(M+\bar M)^2$ is reliable, 
in particular $Re(M)<O(1)$, these results 
are valid.
Similar analyses can be done for the polynomial 
K\"ahler potential.
However, it is not clear that the expansion of the K\"ahler potential 
$K(M,\bar M) = \frac{1}{2}(M+\bar M^2) + \cdots $ is reliable 
for $Re(M)\geq O(1)$.
Thus, in the next subsection we will 
perform the same analysis by assuming 
$K=-\ln(M+\bar M)$ as a trial.
That is an example of K\"ahler potentials which have 
the behavior opposite to the canonical form at large $M$.

\subsubsection{The case with $K=-\ln(M+\bar M)$}

Here the same analysis as subsection 3.2.1 will be done 
with the assumption $K=-\ln(M+\bar M)$.
In this case, it is convenient to define 
\begin{equation}
m' \equiv (M+\bar M)\sigma\Delta +1 .
\end{equation}
Using this variable, we can write the scalar potential 
(\ref{scalar-V-M}) 
\begin{equation}
V = \frac{\sigma \Delta}{(S+\bar S)(m'-1)}e^{-\Delta (S + \bar S)+2}
e^{-m'} (m'^2+g(S+\bar S))  .
\end{equation}
The solutions of the stationary condition 
$\frac{\partial V }{\partial m'}=0$ are obtained 
\begin{equation}
m'=0, \qquad m'=1\pm \sqrt{-g(S + \bar S)-1} .
\end{equation}
The latter solution is allowed only if 
\begin{equation}
g(S+\bar S) <-1.
\end{equation}

For $\sigma  < 0$, the region with $Re(M) > 0$ 
corresponds to $m'<1$.
In this case, the second derivative of the scalar potential,
$\frac{\partial^2 V}{\partial m'^2}$ is positive at $m'=0$ 
if 
\begin{equation}
g(S + \bar S) > -2.
\end{equation}
This is always satisfied by the definition (\ref{def-g}) if 
$(S + \bar S) >0$.
At $m'=1-\sqrt{-g(S + \bar S)-1}$, we have 
$\frac{\partial^2 V}{\partial m'^2} > 0$  if 
\begin{equation}
g(S + \bar S) < -2.
\end{equation}
This is never satisfied by the definition (\ref{def-g}) if 
$(S + \bar S) >0$.
Fig. 3 shows $v=-\frac{e^{-m'}}{m'-1} [{m'}^2+g(S+\bar S)]$ 
for $g(S+\bar S)=-1.5$ and $-3$, respectively.
The scalar potential has a singularity at $m'=1$, 
which comes from the singularity of the K\"ahler potential 
at $M=0$.
However, at the vicinity of $M=0$ the K\"ahler potential
$K(M,\bar M)= \frac{1}{2}(M+ \bar M)^2$ as studied in the 
previous subsection is rather reliable than the 
K\"ahler potential $-\ln(M + \bar M)$.

\begin{figure}
\epsfxsize=0.7\textwidth
\centerline{\epsfbox{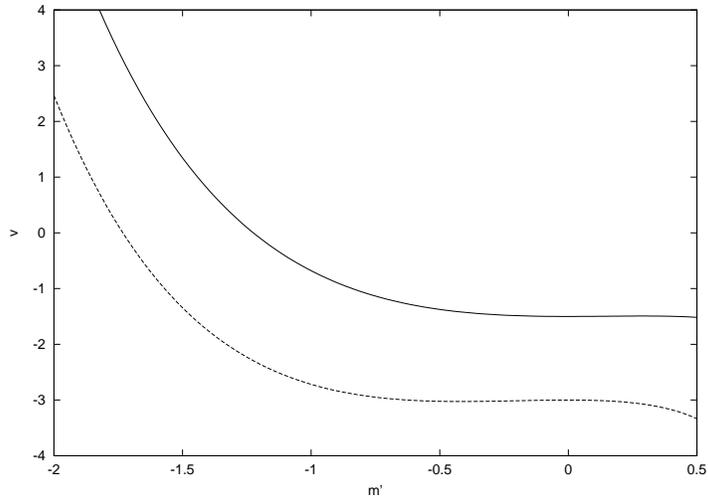}}
\caption{The upper and lower lines show 
$v=-\frac{e^{-m'}}{m'-1} [{m'}^2+g(S+\bar S)]$ 
for $g(S+\bar S)=-1.5$ and $-3$, respectively.}
\label{fig3}
\end{figure}

For $\sigma >  0$, the region with $Re(M) > 0$ 
corresponds to $m'>1$.
However, the second derivative of the scalar potential,
$\frac{\partial^2 V}{\partial m'^2}$ is always negative at 
$m'=1+ \sqrt{-g(S + \bar S)-1}$.
This situation is the same as the problem 
of the dilaton stabilization by the single 
gaugino condensation as seen in Section 2.

Thus, the model with $\sigma  < 0$ is interesting 
for $K(M,\bar M) = -\ln(M + \bar M)$, 
that is, the positive exponent of $M$ in the superpotential 
is useful.
The solutions $m'=0$ and $1 - \sqrt{-g(S+\bar S)-1}$ 
corresponds to 
\begin{equation}
Re(M) = \frac{-1}{\sigma \Delta}, \qquad 
\frac{-\sqrt{-g(S+ \bar S)-1}}{\sigma \Delta},
\end{equation}
respectively.

Assuming the K\"ahler potential 
$K(M,\bar M)=\frac{1}{2}(M + \bar M)^2$ and $-\ln(M + \bar M)$, 
we have shown that the VEV of $Re(M)$ can be stabilized 
with the VEV of $S$ fixed.
The former case implies that the canonical K\"ahler potential 
is important for the stabilization of the twisted moduli.
Such analysis can be extended into the case with 
polynomial K\"ahler potential $K(M,\bar M)$.
On the other hand, the latter case with 
$K(M,\bar M)= -\ln(M + \bar M)$ shows that even 
with the logarithmic K\"ahler potential, the positive 
exponent in the non-perturbative superpotential is 
useful to stabilize the VEV of twisted moduli fields.
It is speculative whether really 
$K(M,\bar M) =-\ln(M+\bar M)$ for large $M$, but that is 
an example of K\"ahler potentials which have the behavior 
opposite to the canonical form at large $M$.
For other forms of K\"ahler potential, the analysis can be 
extended.
The key-point in the stabilization of twisted moduli is that 
the polynomial form of K\"ahler potential is useful and 
the positive exponent of $M$ in the superpotential is 
helpful.
These aspects differ from the dilaton stabilization.
The positive exponent of the dilaton field in the 
non-perturbative superpotential corresponds to the 
asymptotically non-free case \footnote{See Ref.~\cite{Burgess:1997pj} 
for the dilaton
stabilization in the asymptotically non-free case.}.

Of course, the VEV of $S$ is not stabilized in the 
case with the single gaugino condensation which discussed 
in this subsection.
In order to study the stabilization of $S$ and $M$ at the same time, 
we will consider the double gaugino condensations 
in the following subsections.

\subsection{Mirage model}

Here we consider the superpotential generated from the 
double gaugino condensations, i.e., the racetrack model,
\begin{equation}
W = d_1 e^{-\Delta_1 (S + \sigma_1 M)} 
+ d_2 e^{-\Delta_2 (S + \sigma_2 M)} .
\end{equation}

The mirage unification can occur in the case that 
$\sigma^{MSSM}_a$ for the MSSM are proportional to the 
one-loop beta-function coefficients $b^{MSSM}_a$.
Here we consider a specific type of gaugino condensation models 
that $\sigma_a$ for the double gaugino condensations are 
proportional to their one-loop beta-function coefficients, 
that is, we can write 
\begin{equation}
\Delta_a \sigma_a  = C,
\end{equation}
where $C$ is common for the double gaugino condensations for $a=1,2$.
Then, the superpotential can be written as 
\begin{equation}
W = e^{-CM} \tilde W, \qquad 
\tilde W = (d_1 e^{-\Delta_1 S} + d_2 e^{-\Delta_2 S}) .
\end{equation}
The corresponding scalar potential is written as 
\begin{eqnarray}
V &=& \frac{e^{K(M,\bar M)-C(M+\bar M)}}{S + \bar S}
[(K^{-1})^M_{\bar M}|\frac{\partial K(M,\bar M)}{\partial M}-C|^2 
|\tilde W|^2 \nonumber \\
&+& |(S+\bar S)\tilde W_S -\tilde W|^2 - 3|\tilde W|^2].
\end{eqnarray}
The racetrack solution (\ref{RT-sol-1},\ref{RT-sol-2}) corresponding to 
$(S+\bar S)\tilde W_S -\tilde W =0$ is still a solution of 
$\frac{\partial V}{\partial S} =0$ for the present scalar potential.
Here we restrict ourselves to this solution and 
the VEV of $S$ itself is obtained by Eqs.(\ref{RT-sol-1},\ref{RT-sol-2}).
The analysis on the scalar potential for the twisted moduli 
is almost the same as what has been done in section 3.2.
The present case corresponds to the case with 
$g(S + \bar S)=-3$ and $\Delta \sigma = C$.

To be concrete, we again use the assumption of 
$K(M,\bar M)$ as the canonical form and logarithmic form.
First, in the case with $K(M,\bar M)= \frac{1}{2}(M + \bar M)^2$ 
the solutions of $\frac{\partial V}{\partial m} =0$ are 
obtained as
\begin{equation}
Re(M) = \frac{C}{2},\quad \frac{C\pm 1}{2}.
\end{equation}
For the former solution, we have 
$\frac{\partial^2 V}{\partial m^2} < 0$ 
because of $g(S + \bar S)=-3$.
If there are additional contributions increasing the value of 
$g(S + \bar S)$, this solution could be a local minimum.
On the other hand, for the latter solution $2Re(M)= C \pm 1$, 
we have $\frac{\partial^2 V}{\partial m^2} > 0$ 
as well as $\frac{\partial^2 V}{\partial m \partial S} > 0$.
At this point, the $F$-component of $M$ is 
obtained as 
\begin{equation}
|F_M| = \frac{1}{\sqrt{S + \bar S}} e^{\frac{1}{4}(1-C^2)}|\tilde W|.
\end{equation}

Similarly, we can analyse the potential minima for the 
assumed K\"ahler potential 
$K(M,\bar M) = - \ln (M + \bar M)$.
We are interested in the case with $C <0$.
The solutions of $\frac{\partial V}{\partial m'} =0$ are 
obtained as
\begin{equation}
Re(M) = -\frac{1}{2C}, \quad 
\pm \frac{1}{\sqrt{2}C} .
\end{equation}
For the former solution,  $Re(M)=-\frac{1}{2C}$,  
we have $\frac{\partial^2 V}{\partial m'^2} < 0$ 
because of $g(S + \bar S) =-3$.
Additional contributions increasing $g(S + \bar S)$ 
might make this point a local minimum.
For the solution $Re(M) = -\frac{1}{\sqrt{2}C}$,
we have $\frac{\partial^2 V}{\partial m'^2} > 0$ 
as well as $\frac{\partial^2 V}{\partial m' \partial S} > 0$.
At this point, the $F$-component of $M$ does not vanish.
Furthermore, stabilized values must satisfy the constraint 
$Re(f_a) = Re(S) + \sigma_a Re(M) > 0 $.
For the above solution $Re(M)=-\frac{1}{\sqrt 2C}$, 
we can write
\begin{equation}
Re(f_a) = Re(S) -\frac{1}{\sqrt2 \Delta_a}.
\end{equation}
Thus, the stabilized value of $S$ (\ref{RT-sol-2}) must satisfy 
$Re(S) > \frac{1}{\sqrt2 \Delta_a}$.
Recall that the natural order of $S$ is of $O(\frac{1}{\Delta})$ 
unless $\Delta_a$ are close each other or 
$\ln\frac{\Delta_1 d_1}{\Delta_2 d_2}$ is large.

\subsection{Generic racetrack model}

In the previous section, we considered the specific 
racetrack model, i.e. 
$\sigma_1 \Delta_1 = \sigma_2 \Delta_2$.
For generic case $\sigma_1 \Delta_1 \neq  \sigma_2 \Delta_2$, 
analyses become complicated.
Here we give a comment on such generic case.

As solutions of $\partial V /\partial S =0$, 
we again concentrate on the solution (\ref{sol-1}).
That leads to the following equations,
\begin{equation}
\ln \frac{(2\Delta_1 Re(S)+1)d_1}{(2\Delta_2 Re(S)+1)d_2} = 
(\Delta_1 - \Delta_2)Re(S) + (\Delta_1 \sigma_1 - \Delta_2 \sigma_2)
Re(M) ,
\label{RT-sol-3}
\end{equation}
\begin{equation}
(\Delta_1 - \Delta_2)Im(S) + (\Delta_1 \sigma_1 - \Delta_2 \sigma_2)
Im(M) = (2n+1)\pi .
\end{equation}
Furthermore, if $\Delta Re(S) \gg 1$, we obtain 
\begin{equation}
Re(S) = \frac{
(\Delta_1 \sigma_1 - \Delta_2 \sigma_2)}
{\Delta_2 - \Delta_1}Re(M) + \frac{1}{\Delta_1 - \Delta_2}
\ln \frac{\Delta_1 d_1}{\Delta_2 d_2} .
\label{RT-sol-4}
\end{equation}
The second term in the right hand side is the same as 
Eq.(\ref{RT-sol-2}).
The first term is a new contribution from $M$.
When $\Delta_1 \sigma_1 = \Delta_2 \sigma_2$, 
the first term vanishes and that is consistent with 
the subsection 3.3.
However, if $\Delta_1 \sigma_1 \neq \Delta_2 \sigma_2$, 
the VEV of $Re(M)$ corresponds  effectively to  
a large difference of $d_a$ in Eq.(\ref{RT-sol-2}) as 
seen by replacing $d_a \rightarrow d_a e^{-\Delta_a \sigma_aRe(M)}$.
Thus, the value of $Re(M)$ is important to 
the stabilized value of $Re(S)$.

Suppose that the VEV of $Re(M)$ is also stabilized 
by the following equation similar to Eq.(\ref{sol-1})
\begin{equation}
\frac{\partial K(M,\bar M)}{\partial M}W + W_M = 0.
\end{equation}
Combined with Eq.(\ref{RT-sol-3}), for $\Delta Re(S) \gg 1$, 
we obtain 
\begin{equation}
\frac{\partial K}{\partial M}  
= \frac{\Delta_1 \Delta_2(\sigma_1 - \sigma_2)}{(\Delta_2 - \Delta_1)} .
\end{equation}
For example, that leads to 
\begin{equation}
Re(M) = \frac{\Delta_1 \Delta_2(\sigma_1 - \sigma_2)}
{2(\Delta_2 - \Delta_1)},
\end{equation}
for $K(M,\bar M) = \frac{1}{2}(M +\bar M)^2$, and 
\begin{equation}
Re(M) = \frac{\Delta_2 - \Delta_1}
{2\Delta_1\Delta_2(\sigma_1-\sigma_2)} ,
\end{equation}
for $K(M,\bar M) =-\ln (M + \bar M)$.
In the former (latter) case, the value of $Re(M)$ is enhanced 
(suppressed) for fine-tuning $\Delta_1 \approx \Delta_2$, 
while it is suppressed (enhanced) for 
fine-tuning $\sigma_1 \approx \sigma_2$.
Eq.(\ref{RT-sol-4}) becomes 
\begin{equation}
Re(S) = (-\sigma_1 + \frac{2}{\Delta_1}Re(M))Re(M) + 
\frac{1}{\Delta_1 - \Delta_2}
\ln \frac{\Delta_1 d_1}{\Delta_2 d_2}
\end{equation}
for $K(M,\bar M) = \frac{1}{2}(M +\bar M)^2$, and 
\begin{eqnarray}
Re(S) 
&=& -\sigma_1 Re(M) + \frac{1}{2\Delta_1}+ 
\frac{1}{\Delta_1 - \Delta_2}
\ln \frac{\Delta_1 d_1}{\Delta_2 d_2} ,
\end{eqnarray}
for $K(M,\bar M) =-\ln (M + \bar M)$.
For the latter case, the first term in the right hand side 
would be important when $Re(M)$ is enhanced by fine-tuning 
$\sigma_1 \approx \sigma_2$.
Thus, the value of $Re(M)$ has the interesting effect 
on the stabilized value of $Re(S)$.

\section{Conclusion}

We have studied stabilization of the dilaton and twisted moduli 
by assuming the canonical and logarithmic forms for  
the K\"ahler potential of the twisted moduli field.
The canonical K\"ahler potential plays a role in 
the stabilization of the twisted moduli.
This analysis can be extended into the case with 
polynomial K\"ahler potential.
On the other hand, even with the logarithmic K\"ahler potential, 
the positive exponent of the twisted moduli field 
in the non-perturbative superpotential is 
significant.
The logarithmic form was used as an example of K\"ahler potentials 
which have different behavior from the canonical form.
That suggests that even for such case the positive 
exponent of the twisted moduli fields in the superpotentail 
would be helpful.
These aspects are different from the dilaton stabilization.

Simialrly, in the models that gauge kinetic functions depend 
linearly on two or more moduli fields, the positive exponent 
of those fields in the superpotential might be helpful 
for the moduli stabilization.

Also we have considered the specific racetrack model 
with $\sigma_1 \Delta_1 = \sigma_2 \Delta_2$ 
in order to discuss stabilization of the 
dilaton and twisted moduli at the same time.
In generic case, the VEV of $M$ affects 
the stabilized value of the dilaton VEV. 
This point is also important in the stabilization of 
the twisted moduli fields.

The knowledge on the K\"ahler potential of 
the twisted moduli field is necessary to investigate 
numerically reliable results.
We have not taken into account $D$-terms.
Inclusion of $D$-terms would be studied elsewhere.

The models have been studied lead to the negative cosmological 
constant.
That is a common problem as the dilaton stabilization.
The vanishing cosmological constant could be realized by 
the models with more gaugino condensations \cite{deCarlos:1992da}, 
non-perturbative K\"ahler potential \cite{Casas:1996zi} 
or R-symmetry \cite{Izawa:1998dv}.

\section*{Acknowledgment}

The authors would like to thank 
Kiwoon Choi, Yoshiharu Kawamura and Hiroaki Nakano for useful discussions.
T.~K. is supported in part by the Grant-in-Aid for 
Scientific Research from Ministry of Education, Science, 
Sports and Culture of Japan (\#14540256).

\end{document}